\documentclass[10pt,letterpaper]{article}
\usepackage[top=0.85in,left=2.75in,footskip=0.75in]{geometry}

\usepackage{amsmath,amssymb}

\usepackage{changepage}

\usepackage[utf8x]{inputenc}

\usepackage{textcomp,marvosym}

\usepackage{cite}

\usepackage{nameref,hyperref}

\usepackage{microtype}
\DisableLigatures[f]{encoding = *, family = * }

\usepackage[table]{xcolor}

\usepackage{array}

\newcolumntype{+}{!{\vrule width 2pt}}

\newlength\savedwidth

\raggedright
\usepackage[aboveskip=1pt,labelfont=bf,labelsep=period,justification=raggedright,singlelinecheck=off]{caption}

\makeatletter
\renewcommand{\@biblabel}[1]{\quad#1.}
\makeatother

\usepackage{lastpage,fancyhdr,graphicx}
\usepackage{epstopdf}
\fancyheadoffset[L]{2.25in}
\fancyfootoffset[L]{2.25in}
\begin{document}
\vspace*{0.2in}

% Title must be 250 characters or less.
\begin{flushleft}
{\Large
\textbf\newline{Predicting Zip Code-Level Vaccine Hesitancy in US Metropolitan Areas Using Machine Learning Models on Public Tweets} % Please use "sentence case" for title and headings (capitalize only the first word in a title (or heading), the first word in a subtitle (or subheading), and any proper nouns).
}
\newline
% Insert author names, affiliations and corresponding author email (do not include titles, positions, or degrees).
\\
Sara Melotte\textsuperscript{1},
Mayank Kejriwal\textsuperscript{1,*}
%Name3 Surname\textsuperscript{2,3\textcurrency},
%Name4 Surname\textsuperscript{2},
%Name5 Surname\textsuperscript{2\ddag},
%Name6 Surname\textsuperscript{2\ddag},
%Name7 Surname\textsuperscript{1,2,3*},
%with the Lorem Ipsum Consortium\textsuperscript{\textpilcrow}
\\
\bigskip
\textbf{1} Information Sciences Institute, University of Southern California, Marina del Rey, CA, United States
\\
\bigskip

% Insert additional author notes using the symbols described below. Insert symbol callouts after author names as necessary.
% 
% Remove or comment out the author notes below if they aren't used.
%
% Primary Equal Contribution Note
%\Yinyang These authors contributed equally to this work.

% Additional Equal Contribution Note
% Also use this double-dagger symbol for special authorship notes, such as senior authorship.
%\ddag These authors also contributed equally to this work.

% Current address notes
%\textcurrency Current Address: Dept/Program/Center, Institution Name, City, State, Country % change symbol to "\textcurrency a" if more than one current address note
% \textcurrency b Insert second current address 
% \textcurrency c Insert third current address

% Deceased author note
%\dag Deceased

% Group/Consortium Author Note
%\textpilcrow Membership list can be found in the Acknowledgments section.

% Use the asterisk to denote corresponding authorship and provide email address in note below.
* kejriwal@isi.edu

\end{flushleft}
% Please keep the abstract below 300 words
\section*{Abstract}
Although the recent rise and uptake of COVID-19 vaccines in the United States has been encouraging, there continues to be significant vaccine hesitancy in various geographic and demographic clusters of the adult population. Surveys, such as the one conducted by Gallup over the past year, can be useful in determining vaccine hesitancy, but can be expensive to conduct and do not provide real-time data. At the same time, the advent of social media suggests that it may be possible to get vaccine hesitancy signals at an aggregate level (such as at the level of zip codes) by using machine learning models and socioeconomic (and other) features from publicly available sources. It is an open question at present whether such an endeavor is feasible, and how it compares to baselines that only use constant priors. To our knowledge, a proper methodology and evaluation results using real data has also not been presented. In this article, we present such a  methodology and experimental study, using publicly available Twitter data collected over the last year. Our goal is not to devise novel machine learning algorithms, but to evaluate existing and established models in a comparative framework. We show that the best models significantly outperform constant priors, and can be set up using open-source tools.

% Please keep the Author Summary between 150 and 200 words
% Use first person. PLOS ONE authors please skip this step. 
% Author Summary not valid for PLOS ONE submissions.   
%\section*{Author summary}
%Lorem ipsum dolor sit amet, consectetur adipiscing elit. Curabitur eget porta erat. Morbi consectetur est vel gravida pretium. Suspendisse ut dui eu ante cursus gravida non sed sem. Nullam sapien tellus, commodo id velit id, eleifend volutpat quam. Phasellus mauris velit, dapibus finibus elementum vel, pulvinar non tellus. Nunc pellentesque pretium diam, quis maximus dolor faucibus id. Nunc convallis sodales ante, ut ullamcorper est egestas vitae. Nam sit amet enim ultrices, ultrices elit pulvinar, volutpat risus.

%\linenumbers

% Use "Eq" instead of "Equation" for equation citations.
\section*{Background}
Although more people continue to be vaccinated against COVID-19 in the United States and many other nations with each passing week, significant vaccine hesitancy continues to persist \cite{vacc1}, \cite{vacc2}. Vaccine hesitancy in the US has complex drivers, not all of which are well understood, especially among under-served segments of the population \cite{covidHes1}, \cite{covidHes2}. Even prior to COVID-19, vaccine hesitancy against influenza (among others) was non-trivial \cite{vaccHes1}. In the early days of COVID-19, conspiracy theories about the vaccines (as well as other COVID-related issues) had a significant footprint on social media \cite{misinfo1}, \cite{misinfo2}. Such theories, as well as other sources of disinformation and misinformation, can not only go viral on social media but lead to real-world public health consequences by sowing doubt and mistrust among people who are otherwise `on the fence' about the benefits and efficacy of vaccines. Unfortunately, it can be challenging to detect vaccine hesitancy, whether severe or moderate, in near-real time. Surveys, such as the one conducted by Gallup \cite{gallupSurvey}, released early enough, are expensive and still subject to lags. Given the frenetic pace of digital communication and social media virality, more real-time (and inexpensive) detection of vaccine hesitancy at privacy-preserving spatial granularity, such as at the county or even zip code-level, can significantly increase public utility. Even in a post-COVID era, such a system, if sufficiently generalized, can help detect and address vaccine hesitancy (for influenza and other diseases) before it becomes entrenched in a particular region. 

Recent advances in natural language processing (NLP) and social media analysis have been quite impressive, especially due to improvements in deep neural networks and language `representation' learning \cite{bert}, \cite{word2vec}, \cite{fastText}. For example, so-called `word embedding' algorithms, which have been trained on large quantities of text in corpora such as Wikipedia and Google News, learn a real-valued vector representation for each word \cite{word2vec}. In the vector space, the semantics of words are captured in an intuitive way. Even early word embedding algorithms were capable of analogical reasoning (e.g., the vector obtained from the operation $\vec{King}-\vec{Man}+\vec{Woman}$ was found to be closest to the vector $\vec{Queen}$). What is impressive about these embeddings is that the neural network is able to learn such patterns simply by `reading' the text rather than requiring explicitly labeled data from humans. Modern representation learning algorithms, proposed in the last 5-7 years are now capable of embedding sentences (including tweets), which can be used in machine learning models to  make task-specific predictions, given a small set of tweets labeled by hand or using survey data \cite{bert}, \cite{gpt3}. 

In this article, we consider a task that is especially relevant in digital health; namely, the task of predicting vaccine hesitancy using public social media data (with Twitter as the main example), at least in metropolitan areas which are known for high tweeting activity, and where users tend to enable the location facility on their phone compared to less urban milieus \cite{malik2015a}. We are not looking to predict vaccine hesitancy at an individual level, both because it violates privacy but also because we cannot evaluate the accuracy of such predictions without polling the individual. Rather, we seek a way of predicting vaccine hesitancy at the \emph{zip code-level} that uses the text in public, geolocated tweets, but that does not identify or isolate user data of any kind. One advantage of making predictions at the zip code-level is that predictions can be validated using independent survey data, such as the recent poll conducted by Gallup where a question was posed to individuals that is directly pertinent to the individual's vaccine hesitancy. As detailed subsequently, by averaging responses of individuals within a given geographical zone (such as a zip code-demarcated region), we can obtain a real-valued vaccine hesitancy estimate for that zip code. 

With the appropriate performance metrics in place (that are subsequently introduced and discussed in the \emph{Evaluation Methodology and Metrics} sub-section within  \emph{Materials and Methods}), one obvious set of baselines that does not rely on any social media data is to just predict a constant-valued vaccine hesitancy estimate (such as 0.5 or 1.0, or even just the average observed in survey data). We show that our machine learning models outperform such measures, with our best machine learning models  achieving a ten percent relative improvement over the best constant-valued baseline (which itself relies on `privileged' information, namely the mean vaccine hesitancy observed in the survey). Our models are practical and guided by real-world intuitions e.g., we not only consider the text and hashtags directly observed in geolocated tweets, but we also consider the use of NLP software for sentiment analysis, as well as external data sources and features (such as the number of hospitals or scientific establishments in a zip code) that may be relevant to estimating vaccine hesitancy. Rather than propose a `winning' model, we provide a full comparison of the different models and features to better understand the performance differences and tradeoffs.  

The rest of this article proceeds as follows. We begin with a comprehensive description of the \emph{Materials and Methods} used in our study, including details on the Twitter-based dataset and its collection, data preprocessing, feature extraction, vaccine hesitancy ground-truth (obtained from independent survey data), evaluation methodology and metrics, and predictive models and baselines. We only use open-source packages and public data in this study, to enable maximal replicability and minimize cost. We follow the discussion on materials and methods with a detailed summary of our key results, followed by qualitative discussion and error analysis. Finally, we conclude the work with a brief primer on possible future directions and caveats.

\section*{Materials and Methods}

\subsection*{Twitter Dataset}
We sample tweets related to the COVID-19 pandemic from the 9 most populous metropolitan areas in the United States \cite{footnote1}. In order of decreasing order of population size, these are: New York, Los Angeles, Chicago, Houston, Phoenix, Philadelphia, San Antonio, San Diego, and Dallas. Our sampled tweets come from the GeoCOV19Tweets dataset \cite{geocov19}, which collects geo-tagged tweets related to the COVID-19 pandemic daily based on certain COVID-specific keywords and hashtags. The dataset also publishes corresponding sentiment scores for each tweet. Only the tweet ID and sentiment scores are published online, in keeping with Twitter's terms and conditions. 
%We provide more details in the subsequent \emph{Features from External Sources} section, and also in the \emph{SI Appendix}. 

In previous work \cite{dataSource}, we `hydrated' (or retrieved, in full) tweets from the GeoCOV19Tweets dataset from March 20, 2020 through December 1, 2020 (255 days). The period from October 27, 2020 - October 28, 2020 was skipped because no sentiment scores were published at the time of our sampling. From the hydrated tweet object (a data structure described extensively in Twitter's developer documentation \cite{twitterdev}), we extracted each tweet's `coordinates' object to ensure that an exact location for the tweet can be determined. These coordinates were then used to filter the tweets by metropolitan area. To do so, we determined whether the tweet's coordinates of origin fell inside a manually-drawn `bounding box' demarcating each of the metropolitan areas listed above.  If falling within a bounding box, the tweet was marked as originating from the corresponding metropolitan area. The coordinates of these bounding boxes can be found in our previous work, along with detailed descriptions of our collection methodology \cite{dataSource}.

In this study, we re-hydrate this collection of tweets using the \emph{twarc}  library to save the tweet's full text and tweet ID \cite{twarc}. After removing any archived tweets or tweets whose `coordinates' object is no longer available, we retain 45,899 tweets. We also collect every tweet's zip code of origin using an Application Programming Interface (API) provided by Geocodio \cite{geocodio}. Geocodio is a company that was founded in 2014 specifically to provide human-readable location information (such as state, city and country) by taking as input a pair of latitude-longitude coordinates.

We also eliminate all zip codes with fewer than 10 tweets, removing 4,799 tweets and 1,321 zip codes. We then merge the data with the zip code-level attributes described subsequently in \emph{Features from External Sources} and remove remaining rows with null values. This leaves us with a total of 29,458 tweets, each of which belongs to one of 493 unique zip codes across the 9 metropolitan areas listed above. None of the 29,458 tweets is a retweet. Table \ref{tab1} shows the overall dataset statistics, including the number of hashtags before and after text preprocessing (explained subsequently).

\begin{table}[ht]

\caption{\emph{Overall dataset statistics per metropolitan area. We report hashtag data both before and after text preprocessing.}}
          \begin{center} \footnotesize 
          \begin{tabular}{|p{0.8in}|p{0.5in}|p{0.45in}|p{0.55in}|p{0.5in}|p{0.5in}|p{0.65in}|} \hline
          {\bf Metropolitan Area}	&  {\bf Num. Zip Codes} &	{\bf Num. Tweets} &	{\bf Avg. Tweets per Zip Code}	& {\bf Num. Hashtags (before)} & {\bf Num. Hashtags (after)} &  {\bf Num. Unique Hashtags (after)} \\ \hline
New York	 & 152	 & 12,612	 & 82.974	 & 41,232	 & 40,419 & 	10,764 \\ \hline
Los Angeles & 	148 & 	10,532 & 	71.162 & 	37,030 & 	36,507 & 	12,422 \\ \hline
Chicago	 & 43	 & 1,544	 & 35.907	 & 3,857	 & 3,792	 & 1,891 \\ \hline
Houston & 	36	 & 1,529 & 	42.472 & 	5,557	 & 5,505 & 	2,526 \\ \hline
San Diego	 & 27	 & 1,061 & 	39.296 & 	3,019	 & 2,980	 & 1,769 \\ \hline
Philadelphia & 	24	 & 817 & 	34.042 & 	2,753	 & 2,727	 & 1,349 \\ \hline
Dallas & 	33 & 	664 & 	20.121 & 	2,250	 & 2,231 & 	1,211 \\ \hline
Phoenix & 	20 & 	496 & 	24.80 & 	1,576 & 	1,563 & 	1,003 \\ \hline
San Antonio & 	10 & 203	 & 20.30  & 	765 & 	756	 & 340 \\ \hline
Total & 	493 & 	29,458 & N/A & 98,039 & 96,480 & N/A \\ \hline

                                    \end{tabular}
                \end{center}\label{tab1}
                \end{table}

\subsection*{Text Preprocessing}
Using a hydrated tweet's full text, we tokenize, make lowercase, and remove mentions using \emph{TweetTokenizer} \cite{tweettokenizer} from the Natural Language Toolkit (NLTK) package \cite{nltk}. NLTK is a leading package in the NLP community that uses Python programs to work with human language data. We also remove URLs, stop words, tokens less than or equal to 1 character in length, and any characters other than letters (including the \# symbol and emojis). We use NLTK's standard set of English stop words \cite{nltkstopwords} (such as `the', `a' etc.) but remove the words \emph{not, no, nor, very,} and \emph{most} from this pre-determined set, as these are assumed to be relevant for making more accurate vaccine hesitancy predictions. We then lemmatize all tokens using \emph{WordNetLemmatizer} \cite{wordnetlemmatizer}. Note that after our text preprocessing steps, the hashtags ``covid19'', ``covid'', ``Covid19'', and ``covid-19'' (for instance) all result in the same token, and any hashtags consisting of numbers or single characters only (e.g. ``\#2020'' or ``\#K'') are eliminated. The processed tweets are then embedded using the fastText word embedding (briefly introduced in \emph{Background})  model, which was released by Facebook AI research and contains word vectors trained on English Wikipedia data \cite{fastTextModel}, as explained further in \emph{Predictive Models and Features}. 

The number of hashtags in the tweets before text preprocessing in Table \ref{tab1} is computed by summing the occurrences of \# in the full text. After text preprocessing, when the hashtags and text are well-separated and more easily analyzed, we count the number of times a token begins with the \# symbol. 

Note that, in this study, we avoid using the `hashtags' object \cite{objectmodel} returned directly from the Twitter API for several reasons. First, the object appears to already have certain filters applied such that number-only strings (e.g. \#2020) are eliminated. Even though our text preprocessing steps do so as well, as mentioned above, the `hashtags' object does not accurately represent the number of hashtags in the original tweet, especially as it fails to accurately count hashtags in a continuous string. For example, in a tweet from New York containing ``...\#corona\#coronavirus\#quarantine\#quarantinelife\#washyourhands...'', the `hashtags' object returns an empty array. To maintain control over text preprocessing and feature extraction, we therefore exclusively use and count hashtags from the full text field in the tweet data dictionary \cite{objectmodel} returned from the Twitter API.

\subsection*{Features from External Sources}

As mentioned earlier in \emph{Twitter Dataset}, we obtain the zip code of origin for every tweet included in our dataset, resulting in a total of 493 unique zip codes. For all unique zip codes, we get additional zip code-level information from external, publicly available data sources. These zip code-level attributes, which we add as features in our predictive models (see \emph{Predictive Models and Features}), comprise the Zillow Home Value Index (ZHVI) \cite{zhvi} and the numbers of establishments in the educational, healthcare, and professional, scientific, or technical sectors. We incorporate these features because they act as proxies for zip code affluence and resource availability. The sentiment scores, which have tweet-level granularity, are obtained directly from the GeoCOV19Tweets dataset \cite{geocov19}, as noted earlier in \emph{Twitter Dataset}. Importantly, every tweet has its own sentiment score, but tweets originating from the same zip code share identical zip code-level attributes i.e., the zip code-level data are repeated for all tweets belonging to the same zip code.

All features mentioned above are standardized, using the \emph{StandardScaler} function in Python's scikit-learn package \cite{standardscaler}, in the train and test data \emph{separately}, to prevent test data leakage into the training phase. We provide more details about the train and test split in the next section. Table \ref{tab2} summarizes each zip code-level feature as well as sentiment, but we also provide more detailed descriptions of these features in the \emph{SI Appendix}.

\begin{table}[ht]

\caption{\emph{External features collected from publicly available data sources. All but sentiment score are zip code-level attributes, meaning that all tweets with the same zip code will have the same values for these features.}}
          \begin{center} \footnotesize 
          \begin{tabular}{|p{1.2in}|p{3.55in}|} \hline
          {\bf Feature} & {\bf Description} \\ \hline
Sentiment Score &	Score between [-1, 1], where positive values signify positive sentiment and 0 means ``neutral''. The greater the absolute value, the stronger the sentiment. \\ \hline
Zillow Home Value Index (ZHVI) & 	Measure of the typical home value in a zip code (USD), capturing market value in addition to price. \\ \hline
Healthcare and Social Assistance & 	Number of establishments that provide health care and social assistance to individuals. \\ \hline
Educational Services	 & Number of establishments, either privately or publicly owned and either for-profit or not-for-profit, that provide instruction or training. \\ \hline
Professional, Scientific, and Technical Services & 	Number of establishments that provide professional, scientific, and technical services that require a high level of expertise or training. \\ \hline
   \end{tabular}
                \end{center}\label{tab2}
                \end{table}

\subsection*{Vaccine Hesitancy Ground Truth and Train / Test Split}

Throughout this study, we refer to the list of the 493 unique zip codes and their corresponding vaccine hesitancies as the `ground truth'. The vaccine hesitancy values range from 0.0 to 1.0 on continuous scale, and represent the percentage of surveyed residents who have indicated that they are hesitant about receiving a COVID-19 vaccine. 

Specifically, we leverage the 2020-21 vaccine hesitancy data collected through the COVID-19 Gallup survey \cite{gallup2}. Gallup launched a survey on March 13, 2020 that collected people's responses during the COVID-19 pandemic, polling daily random samples of the Gallup Panel (a probability-based, nationally representative panel of U.S. adults). A vaccine hesitancy question was asked starting from July 20, 2020: \emph{If an FDA-approved vaccine to prevent coronavirus/COVID-19 was available right now at no cost, would you agree to be vaccinated?} with the possible response of \emph{Yes} or \emph{No} to investigate a person's willingness to be vaccinated. The proportion of \emph{No} responses in a specific area is defined in this work as vaccine hesitancy. We calculate the proportion of the \emph{No} answers to this question between July 20 and August 30, 2020 at the zip code-level to get a vaccine hesitancy score per zip code.  

The mean vaccine hesitancy across all 493 unique zip codes corresponding to our tweets was calculated to be 0.240. The standard deviation is 0.334, showing that there is significant variance across zip codes, even when limited to the largest metropolitan areas in the US. The minimum and maximum values are 0.00 and 1.00, respectively, indicating complete vaccine acceptance and hesitancy.
Note that these `ground truth' values exist at the zip code-level and measure aggregates. A vaccine hesitancy of 0.5 in a zip code thus means that, on average, half the people in that zip code are vaccine hesitant. While we cannot say anything about an individual tweeter, for predictive modeling purposes, we label a tweet originating from zip code $z$ with the ground-truth vaccine hesitancy score corresponding to zip code $z$. This implies that, if there are $k$ tweets from zip code $z$, then all k tweets are assigned the same `pseudo' vaccine hesitancy label. When reporting the results, for completeness, we report metrics both at the tweet-level (which should be assumed to be weakly labeled using the methodology above), as well as at the zip code-level, which is the true measure of the performance of our system.

We apply a stratified split (using the \emph{StratifiedShuffleSplit} function in the scikit-learn package \cite{sss}) to partition our dataset into train (80\%) and test (20\%) sets based on zip code. This way, both the train and test sets include tweets from all 493 zip codes in approximately equal proportions e.g., 42.8\% of both the train and test sets are tweets from the New York City metropolitan area (since 42.8\% of the overall tweets in our corpus are from New York), and so on. Due to this stratified construction, both the train and test sets include tweets from all 9 metropolitan areas. Overall, there are 23,566 tweets in the train set and 5,892 tweets in the test set. 
%Note that, while the tweet IDs (and vaccine hesitancy labels) in the train and test sets remain identical throughout the study, the text embeddings and the number of external features can vary across the predictive models, depending on the details of each model. Additionally, all predictive models described subsequently are trained on the entire train set and output predictions for the entire test set (5,892 predictions). The only exception to this, as explained in the next section, occurs when computing the 5-fold cross-validation metrics as an additional way to verify that the models are not overfitting to the training data.

\subsection*{Evaluation Methodology and Metrics}

All predictive models and baselines used in this study, and described in the next two sections, are evaluated in two different ways: at the tweet-level and at the zip code-level. The tweet-level evaluation is based on a vaccine hesitancy prediction for every tweet in the test set (a total of 5,892 predictions), while the zip code-level evaluation relies on a single vaccine hesitancy prediction per zip code (a total of 493 predictions). Our predictive models, however, make predictions at the individual tweet-level only. To get 493 zip code-level predictions, we take the average of the tweet-level predictions for every zip code. Formally, for $k$ tweets (in the test set) belonging to zip code $z$ with predicted tweet-level vaccine hesitancies $[\hat{y_1},\ldots,\hat{y_k}]$, the predicted vaccine hesitancy for zip code $z$ is given by the formula:
\begin{equation}
\hat{y_z} = \frac{\sum_{i=1}^k \hat{y_i}}{k}
\end{equation}

Both the tweet-level and zip code-level evaluations use the Root Mean Square Error (RMSE) metric for performance validation and comparison. Given $m$ data points with true (real-valued) vaccine hesitancy labels $[y_1,\ldots,y_m]$, and predicted labels $[\hat{y_1},\ldots,\hat{y_m}]$, the RMSE is given by the formula:
\begin{equation}
RMSE = \sqrt{\frac{\sum_{i=1}^m (y_i - \hat{y_i})^2}{m}}
\end{equation}

For the tweet-level evaluation, each of the $m$ data points represents a tweet, while for the zip code-level evaluations, each data point represents a zip code. Thus, in the tweet-level RMSE score calculation, the `pseudo' tweet-level vaccine hesitancy labels, the assignment of which was described in the previous section, are compared with the tweet-level predictions obtained from the model. In the zip code-level RMSE calculation, similarly, the `ground truth' vaccine hesitancies (obtained directly from Gallup) are compared with the zip code-level predictions obtained from the models. The lower the RMSE score, the lower the predictive error, and the better the model.

We emphasize that, although we train our predictive models at the tweet-level, the tweet-level predictions are auxiliary to obtaining zip code-level predicted vaccine hesitancies. This is because we cannot evaluate model performance at the tweet-level when our `ground truth' values have zip code-level granularity. The vaccine hesitancy labels of the tweets should be thought of as `pseudo' (or `weak') labels, and predictive performance at the level of tweets is only reported for the sake of completeness. The primary goal of this study, as discussed in \emph{Background}, is predicting zip code-level vaccine hesitancies using publicly available individual tweets. 

In addition to computing RMSE scores on the test set, for each predictive model, we also report the average of the 5-fold cross-validated RMSE score (at the tweet-level). To do so, we split the train set into five folds. The first fold contains 4,714 tweets, while the other four folds each contain 4,713 tweets (adding up to 23,566 tweets total, which is the entirety of the train set). Each fold is used as a `test' set once, while the remaining four folds act as the `train' set. Because every fold is used as a `test' set only once, there are five training iterations (corresponding to the number of folds). At every iteration, we obtain one RMSE score that represents the performance of the model trained on four folds and evaluated on the fifth. Over all iterations, therefore, we have 5 RMSE scores of which we report the average in \emph{Results} as a measure of model \emph{robustness} i.e., to further verify that the reported tweet-level RMSE values are not the result of luck on the actual test set (containing 5,892 tweets). 

We do not report the average of the 5-fold cross-validated RMSE scores on the zip code-level because cross-validation is computed during training, and, as mentioned in the previous section, model training is done exclusively at the tweet-level. Thus, we train and cross-validate the predictive models only at the tweet-level, as a test of model robustness.

\subsection*{Predictive Models and Features}

As described in \emph{Text Preprocessing}, we use fastText's word vectors trained on English Wikipedia data to embed tweet text. The resulting vectors are 300-dimensional, and all dimensions are retained throughout the study. We embed the processed full text in three different ways (corresponding to three representations): one which includes \emph{text only} (no hashtags), one that includes \emph{text and hashtags}, and one that contains hashtags only, if available, and text otherwise. We refer to the latter as the \emph{hybrid} representation. For instance (ignoring any text transformations explained in \emph{Text Preprocessing}), the \emph{text only} representation of the tweet ``Be back soon my friends \#corona \#cov19 \#notMyVirus \#quarantinefitness'' would embed only the ``Be back soon my friends'' part. The \emph{text and hashtags} representation would incorporate the entire tweet, and the \emph{hybrid} representation would embed only ``\#corona \#cov19 \#notMyVirus \#quarantinefitness'' because the tweet contains hashtags. Alternatively, the \emph{hybrid} representation of the tweet ``In the hospital not for Corona virus'' would embed the tweet's text because no hashtags are provided. Compared to a representation that uses only hashtags, the hybrid representation is more robust since it uses the text if no hashtags are present.

For each of the three representations described above, we build four predictive models (with a total of 12 models) that incorporate all zip code-level features using support vector regression (SVR), linear regression, and stochastic gradient descent (SGD). For SVR, we implement a radial basis function (RBF) kernel as well as a linear kernel. All of these are established models in the machine learning community; technical details can be found in any standard text \cite{bishop}. Using \emph{SVR with RBF kernel}, we also build three predictive models (one for each representation) that do not incorporate any zip code-level features. We chose \emph{SVR with RBF kernel} because, out of all 12 predictive models just mentioned, it performs the best across all representations, as subsequently shown in \emph{Results}. All predictive models, both including and excluding zip code-level features, are also trained with and without the sentiment feature. We report both scores because sentiment is an external, tweet-level feature not computed or verified by us, but provided by the GeoCOV19Tweets dataset, which is a super-set of our metropolitan data.

Apart from setting the maximum number of iterations in \emph{SVR with linear kernel} to 4,000 and specifying a random state value (42) for \emph{SVR with linear kernel} and SGD, we use default parameters for all 15 predictive models. Recall from the previous section that for each of these models, we compute the RMSE score on the tweet-level and the zip code-level, and we also report the mean of the 5-fold cross-validated RMSE scores (for the tweet-level only).

\subsection*{Baselines}

To evaluate the predictive power of the 15 models described in the previous section, we consider six \emph{constant}-value baselines that predict a single value for each individual tweet (\emph{tweet-level}) and each unique zip code (\emph{zip code-level}). The RMSE scores for the tweet-level baseline predictions are measured with respect to the weakly labeled test set (5,892 tweets), while the errors for the zip code-level baselines are computed with respect to the original zip code-level `ground truth' (493 zip codes). Our baselines do not rely on sentiment, or on any text or zip code-level features. Table \ref{tab3} summarizes the models.

\begin{table}[ht]

\caption{\emph{Description of the six constant-value baseline models, along with the corresponding vaccine hesitancy value predicted for all tweets and zip codes (for that baseline).}}
          \begin{center} \footnotesize 
          \begin{tabular}{|p{1.1in}|p{2.5in}|p{1.2in}|} \hline
          {\bf Baseline System} & {\bf Description} & {\bf Constant-Value Vaccine Hesitancy} \\ \hline
No vaccine hesitancy & 	Predicts that none of the zip codes are vaccine hesitant & 	0.0 \\ \hline
Complete vaccine hesitancy & 	Predicts that all zip codes are completely vaccine hesitant	 & 1.0 \\ \hline
Partial vaccine hesitancy & 	Predicts that all zip codes are partially vaccine hesitant	 & 0.5 \\ \hline
Mean `pseudo' vaccine hesitancy in train set & 	Predicts that all zip codes have a vaccine hesitancy equal to the average vaccine hesitancy of only the train set tweets (23,566 samples)	 & 0.2349 \\ \hline
Mean `pseudo' vaccine hesitancy for all tweets & 	Predicts that all zip codes have a vaccine hesitancy equal to the average vaccine hesitancy of all tweets (29,458 samples) & 	0.2349 \\ \hline
Mean vaccine hesitancy in Gallup ground-truth & 	Predicts that all zip codes have a vaccine hesitancy equal to the average vaccine hesitancy of the ground-truth values (493 samples)	 & 0.2403 \\ \hline

   \end{tabular}
                \end{center}\label{tab3}
                \end{table}

Concerning the last three baselines, both the mean vaccine hesitancy in the train set and the mean vaccine hesitancy for all tweets are weighted by the frequencies of the zip codes in our dataset. Additionally, the baseline relying on information from the ground-truth is highly optimistic because it assumes that this information is known. Even the previous two baselines (at the tweet-level) rely on this information since the pseudo-label relies on the zip code-level label, which is obtained from the ground-truth. In reality, the average vaccine hesitancy of the ground truth, the train set, or the entire dataset will not be available, since that is precisely what we are trying at predict at the zip code-level. In subsequent sections, we refer to these baselines as the `optimistic' baselines in contrast with the first three `realistic' baselines, which assume a manually specified constant value.

\section*{Results}
The RMSE metrics tabulated in Table \ref{tab4} show that all predictive models outperform the best-performing, realistic baseline (``no vaccine hesitancy'' with an RMSE of 0.411) at the zip code-level, although not all perform better than the most optimistic baselines (RMSE of 0.334). Specifically, no model has an RMSE score below 0.334, our optimistic baseline value, except for the \emph{SVR with RBF kernel} models (without sentiment).

\begin{table}

\caption{\scriptsize\emph{The Root Mean Square Error (RMSE) scores at both the tweet-level and zip code-level for all models along with the average of the 5-fold cross-validated RMSE scores (for the predictive models).} Models specified with (*) do not include any zip code-level features. The others include all zip code-level features in addition to text. The RMSE scores for the predictive models with and without sentiment as a feature are reported as ``without sentiment / with sentiment''. Cross-validation is not applicable for the baseline models.}
          \begin{center} \scriptsize 
          \begin{tabular}{|p{0.8in}|p{1.4in}|p{0.7in}|p{0.7in}|p{0.7in}|} \hline
          {\bf Representation}	 &  {\bf Model} &	{\bf Tweet-Level RMSE} &	{\bf Mean of 5-fold Cross-Validation RMSE Scores}	& {\bf Zip Code-level RMSE} \\ \hline
Text Only & SVR (RBF kernel) &  	{\bf 0.206} / 0.211	 &  {\bf 0.209} / 0.214	 &  0.312 / 0.316 \\ \hline
& SVR (RBF kernel)* &  	0.271 / 0.290	 &  0.270 / 0.289 &  	{\bf 0.308} / 0.329 \\ \hline
& SVR (linear kernel) &  	0.297 / 0.297 &  	0.295 / 0.295 &  	0.374 / 0.374 \\ \hline
& Linear regression &  	0.280 / 0.280 &  	0.279 / 0.279 &  	0.336 / 0.336 \\ \hline
& SGD regressor	 &  0.297 / 0.297 &  	0.299 / 0.299 &  	0.343 / 0.343 \\ \hline \hline
Text and Hashtags & SVR (RBF kernel) &  	0.208 / 0.213	 &  0.210 / 0.215 &  	0.315 / 0.318 \\ \hline
& SVR (RBF kernel)* &  	0.269 / 0.291 &  	0.267 / 0.289 &  	0.314 / 0.335 \\ \hline
& SVR (linear kernel) &  	0.297 / 0.297 &  	0.295 / 0.295 &  	0.374 / 0.373 \\ \hline
& Linear regression &  	0.280 / 0.280 &  	0.278 / 0.278 &  	0.335 / 0.355 \\ \hline
& SGD regressor	 &  0.300 / 0.300 &  	0.301 / 0.301 &  	0.343 / 0.343 \\ \hline \hline
Hybrid & SVR (RBF kernel) &  	0.213 / 0.217 &  	0.213 / 0.219 &  	0.319 / 0.320 \\ \hline
& SVR (RBF kernel)* &  	0.287 / 0.287 &  	0.280 / 0.297 &  	0.322 / 0.322 \\ \hline
& SVR (linear kernel) &  	0.303 / 0.303 &  	0.309 / 0.309	 &  0.365 / 0.365 \\ \hline
& Linear regression &  	0.292 / 0.292	 &  0.290 / 0.290	 &  0.334 / 0.334 \\ \hline
& SGD regressor &  	0.307 / 0.307	 &  0.299 / 0.299 &  	0.341 / 0.341 \\ \hline \hline
Constant-Value Baselines & No vac. hes. &  	0.399 &  	N/a &  	0.411 \\ \hline
& Complete vaccine hesitancy &  	0.830 &  	N/A &  	0.830 \\ \hline
& Partial vaccine hesitancy. &  	0.418 &  	N/A &  	0.423 \\ \hline
& Mean `pseudo' vaccine hesitancy in train set &  	0.322 &  	N/A	 &  0.334 \\ \hline
& Mean `pseudo' vaccine hesitancy for all tweets &  	0.322 &  	N/A	 &  0.334 \\ \hline
& Mean vaccine hesitancy in Gallup ground-truth	 &  0.322	 &  N/A &  	0.334 \\ \hline

                                    \end{tabular}
                \end{center}\label{tab4}
                \end{table}

For most predictive models, there is no difference in performance when sentiment is included versus when it is excluded. However, \emph{SVR with RBF kernel} shows several noteworthy improvements when sentiment is omitted as a feature. Without sentiment, \emph{SVR with RBF kernel} is the only model that outperforms even the most optimistic baselines across all 3 representations (\emph{text only, text and hashtags}, and \emph{hybrid}) regardless of whether zip code-level features are included. The only time the \emph{SVR with RBF kernel} model performs worse than our optimistic baselines is when the \emph{text and hashtags} representation is used and when sentiment is included but all zip code-level features are excluded (RMSE of 0.335). However, this is only a 0.1\% increase in RMSE compared with the optimistic baselines. 

We observe the lowest zip code-level RMSE score (0.308) through the \emph{SVR with RBF kernel} model (\emph{text only} representation) when no features other than the text are included (i.e., no zip code-level features and no sentiment). This is a 7.78\% improvement from the optimistic baselines, and a 25.06\% improvement from the best-performing, realistic baseline (``no vaccine hesitancy''). Compared with the ``complete vaccine hesitancy'' baseline, this model shows a 62.89\% improvement, and compared with the ``partial vaccine hesitancy'' baseline, we observe a 27.19\% improvement. When zip code-level features are added, the \emph{text only} representation performs with an RMSE of 0.312, yielding a 6.59\% improvement compared with the optimistic baselines and a 24.09\% improvement compared with the ``no vaccine hesitancy'' baseline. Table \ref{tab4} shows that adding sentiment to these two models in the \emph{text only} representation reduces performance at the zip code-level.

Interestingly, adding sentiment as the only external feature to the text embeddings (no zip code-level attributes) reduces performance greatly for the \emph{text only} and \emph{text and hybrid}  representations using the \emph{SVR with RBF kernel} model. For the former, performance is reduced by 6.38\%, while for the latter, performance decreases by 6.27\%. For the \emph{hybrid} representation, on the other hand, the RMSE score is unaffected.

In addition to the results presented in Table \ref{tab4}, we also built an \emph{SVR with RBF kernel} model that predicts zip code-level vaccine hesitancy based on zip code-level data only (no text), both with and without sentiment. When predicting based on zip code-level data only, this model achieves an RMSE of 0.332, and when sentiment is added, it achieves an RMSE of 0.333. Both outperform the most realistic baseline model (``no vaccine hesitancy'') with 19.22\% and 18.98\% performance improvements, respectively. They also marginally outperform our optimistic baselines, but only with 0.60\% and 0.30\% decreases in RMSE scores, respectively. This suggests that even when relying on only the number of establishments in the healthcare, educational, and professional, scientific, or technology sectors as well as the Zillow Home Value Index (ZHVI), it is possible to predict zip code-level vaccine hesitancy with a marginally lower error compared with constant-value baselines. 

However, as reported above, the addition of tweet text leads to lower error for our zip code-level predictions. Our best-performing model shows a 25.06\% improvement (compared with only 19.22\% when no text is used) over a ``no vaccine hesitancy'' constant-value baseline and a 62.89\% improvement over a ``complete vaccine hesitancy'' constant-value baseline. This indicates that the use of tweet text (particularly in models implementing the \emph{text only} representation and without any external features) is a powerful indicator of zip code-level vaccine hesitancy.  The evidence suggests that tweet-text contains more signal than noise, on average, insofar as predicting vaccine hesitancy is concerned.

\section*{Discussion and Error Analysis}
Although the machine learning-based models presented in the previous sections clearly outperform constant-valued baselines, it is useful to consider the issue from a qualitative lens, both to understand the kinds of tweets that contain either signal or noise, as well as to understand sources of prediction error. To do so, in this section, we sample some tweets from our complete dataset prior to any text preprocessing. We include only the text and hashtags, i.e. we exclude any mentions, hyperlinks, or locations for brevity. Sampled tweets labeled with a vaccine hesitancy of 1.0 (complete vaccine hesitancy) are enumerated below: 

\begin{enumerate}
\item ``Why is COVID-19 mild for some, deadly for others?'' (Chicago)
\item ``Corona quarantine day 19. Trapped on the beach. \#chickendinner \#familytime \#corona \#coronatine \#beachlife'' (San Diego)
\item ``D.C. residents are encouraged to continue practicing social distancing and take actions to help prevent the spread of COVID-19 [PR] Coronavirus Data Update: March 24'' (New York)
\item ``New comfortable and colorful masks for coronavirus protection while you are out and about \#craftfoundry \#covid19masks \#covid19masksforsale \#coronavirusprotection \#coronavirusprotectionmask \#cottonmask'' (Philadelphia)
\end{enumerate}

We also sampled tweets labeled with a predicted vaccine hesitancy of 0.5 by the best model (partial vaccine hesitancy):

\begin{enumerate}
\item ``An emotional week for everyone. Everyone keep fighting. Day by day \#fighting'' (Los Angeles)
\item ``Please no close talking [emoji] practice social distancing \#corona \#codin19 \#thistooshallpass \#seinfeld \#seinfeldmemes \#seinfeldquotes'' (New York)
\item ``good thing the drive-in is open and an easy place for social distancing. \#JurassicPark \#drivein \#movienight \#family'' (San Antonio)
\item ``Covid 19 Testing at First Met! \#firstmetchurch \#covidtesting \#sheilajacksonlee'' (Houston)

\end{enumerate}

Finally, we sampled tweets predicted with a vaccine hesitancy of 0.0 (no vaccine hesitancy):
\begin{enumerate}
\item ``Catch me if you can but don't get burned. \#charmander \#pokemon \#socialdistancing \#corona \#onesie'' (Chicago)
\item ``\#GivingGratitude to our \#frontlineworkers who are working to keep us safe everyday. \#weappreciateyou \#covid19 \#2020 \#amberwavesrealestategroup \#kellerwilliamsdmn \#weareinthistogether \#shelterinplace \#stayhome'' (Dallas)
\item ``This is how you know there's a problem? When is the last time you saw gas prices this low?! \#cov?d19 \#corona'' (Philadelphia)
\item ``I've got this pit in my stomach at whats happening in california and all of the USA during this corona virus pandemic especially as homeless with severe epilepsy makes this worse.'' (Los Angeles)

\end{enumerate} 
These examples illustrate why it's important to measure vaccine hesitancy at an aggregate level, such as zip code, rather than at the individual level using only tweets. Aside from the obvious ethical concerns, it is unlikely that extreme cases of vaccine hesitancy or non-hesitancy will manifest on Twitter for the vast majority of the population (although a full study of this hypothesis merits future research). However, at the zip code-level, the predictions are more promising. 

As presented earlier in Table \ref{tab4}, we showed that the \emph{text only} representation of the \emph{SVR with RBF kernel} model that excludes sentiment and any zip code-level features predicts zip code-level vaccine hesitancies with an RMSE of 0.308. This model, however, tends to overestimate vaccine hesitancy. In 291 zip codes (out of 493), we find that the predicted vaccine hesitancy is greater than the true vaccine hesitancy. In the remaining 202 zip codes, the predicted vaccine hesitancy is lower than the `ground truth' value. For 50 zip codes, the predicted vaccine hesitancy is overestimated by 0.20 or more, and when looking at the difference between the predicted and true vaccine hesitancies (absolute value), we find that the gap is 0.20 or more in 179 zip codes.

For instance, in Philadelphia, 83.33\% of the zip code-level predictions are over-estimates (the highest among all metropolitan areas). In New York, the metropolitan area with the highest proportion of tweets and the highest average number of tweets per zip code in our study, the predicted vaccine hesitancies for 60.53\% of zip codes are overestimated. In fact, in all metropolitan areas except for Phoenix, San Diego, and Dallas, the vaccine hesitancy in more than 50\% of the zip codes is overestimated. In Phoenix, exactly half of all zip code-level vaccine hesitancies are over-estimates. For the remaining metropolitan areas, the percentages of zip codes having overestimated vaccine hesitancies fall between 58.11-62.79\%.

Despite the large proportion of overestimated vaccine hesitancies presented above, we observe that only 16.67\% and 9.21\% of the zip codes in Philadelphia and New York, respectively, have overestimated vaccine hesitancies greater than or equal to 0.20. The metropolitan area with the greatest proportion of zip codes with overestimated vaccine hesitancies by 0.20 or more is San Antonio (30\%), which is also the metropolitan area with the fewest number of tweets. San Diego, on the other hand, has only one zip code (3.70\%) whose vaccine hesitancy is overestimated by 0.20 points or more. 

Lastly, despite the large proportion of overestimates, we observe that Philadelphia, in fact, has the smallest proportion of zip codes (25\%) in which the difference between the predicted and true vaccine hesitancies (absolute value) is greater than or equal to 0.20. Thus, using this metric, we note that our \emph{SVR with RBF kernel} model performed best in Philadelphia. On the contrary, the model is worst performing in Dallas, with 57.58\% of zip codes showing a $\geq 0.20$ difference (absolute value) between the true and predicted vaccine hesitancies. For all other metropolitan areas, the proportion of zip codes whose predicted vaccine hesitancy differs by 0.20 or more points from the true vaccine hesitancy falls between 29.73-50\%. Overall, the results suggest that the methods presented herein should not be used for highly sensitive predictions, but the low gap between the true hesitancy and predicted hesitancy in many zip codes (especially compared to `prior constant' baselines, all of which performed worse than the SVR method) suggests that the method can potentially be used both by social scientists, as well as by digital health experts as an early warning system.

\section*{Conclusion}

With significant (albeit declining) vaccine hesitancy continuing to exist in geographic and demographic segments of the adult population, and the expense of conducting detailed and representative surveys, there is need for cheaper and more real-time measures of predicting vaccine hesitancy without violating privacy. In this article, we explored public social media as a potential source of such information. Without identifying users, our models simply use the text and hashtags in tweets to make a prediction at a zip-code level. We validate the predictions using actual survey data as a gold standard. Experimental results show that such an endeavor is not only feasible, but also reasonably robust i.e., most of our models performed well compared to `constant-valued' baselines even in the presence of irrelevant or non-vaccine related (or even non-COVID related) tweets. Of course, the caveat applies that such a system cannot be a substitute for comprehensive and representative surveys, and it is likely more reliable for urban areas (toward which social media platforms like Twitter tend to be heavily biased in their user-base), especially in primarily-English speaking developed nations like the United States, than for rural areas or for developing nations. However, by serving as a supplementary source of information, such a system may help the public health community to detect clusters of vaccine hesitancy and mitigate it with communication and outreach. We believe that it can serve as a valuable and inexpensive asset in a nation's digital health infrastructure, especially as more of the world comes online, and more people start to engage with social media. In future work, we hope to replicate and extend this work for other cities, countries and languages. 

\paragraph*{S1 Appendix: External Tweet-Level and Zip Code-Level Features}
\label{S1_Appendix}
{\bf Sentiment Score.} We retain the original sentiment scores included in the GeoCOV19Tweets dataset \cite{geocov19} generated using the TextBlob  sentiment analysis tool \cite{textblob}. In this dataset, every tweet is given a continuous value score between [-1, 1], where positive values signify positive sentiment and 0 means ``neutral''. The more positive or negative the value, the stronger the sentiment. Prior to computing these sentiment scores, hashtag symbols (\#), mention symbols (@), URLs, extra spaces, and paragraph breaks were eliminated. Punctuation, emojis, and numbers were included.

{\bf Zillow Home Value Index (ZHVI).} The Zillow Home Value Index (ZHVI)  is a measure of the typical home value for a region; in this case, zip code. It captures monthly changes in Zestimates \cite{zestimates}, which are Zillow's estimated home market values that incorporate house characteristics, market data such as listing prices of comparable of homes and their time on the market, as well as off-market data including tax assessment and public records. It also incorporates market appreciation. In this study, we take the average of the smoothed, seasonally adjusted value in the 35th to 65th percentile range (mid-tier) from January through December 2020.

{\bf Establishments.} Data about the number of establishments per zip code is taken from the 2018 Annual Economic Surveys from the US Census (Table ID CB1800ZBP) \cite{establishments}. We take the ``Health care and social assistance'',  ``Educational services'', and ``Professional, scientific, and technical services'' data, which have the following meaning:

\begin{enumerate}
\item Healthcare and social assistance (sector 62) comprises establishments providing health care and social assistance for individuals \cite{healthcare1} e.g., physician offices, dentists, mental health practitioners, outpatient care centers, ambulance services, etc. \cite{healthcare2}.
\item  Educational services (sector 61) consist of establishments that provide instruction or training in a wide variety of subjects. The sector includes both privately and publicly owned institutions and both for profit and not for profit establishments \cite{education1}  e.g., elementary and secondary schools, colleges, universities, computer training, professional schools, driving schools, etc. \cite{education2}.
\item Professional, scientific, and technical services (sector 54) include establishments that specialize in providing professional, scientific, and technical services that require a high level of expertise or training \cite{professional1} e.g., legal services, notaries, accounting, architectural services, building inspection, engineering services, scientific consulting, research and development, advertising, etc. \cite{professional2}. 
\end{enumerate}
%
%\paragraph*{S1 Table.}
%\label{S1_Table}
%{\bf Lorem ipsum.} Maecenas convallis mauris sit amet sem ultrices gravida. Etiam eget sapien nibh. Sed ac ipsum eget enim egestas ullamcorper nec euismod ligula. Curabitur fringilla pulvinar lectus consectetur pellentesque.

\section*{Acknowledgments}
N/A

%\nolinenumbers

\bibliography{short}
%\bibliographystyle{plain}
% Either type in your references using
% \begin{thebibliography}{}
% \bibitem{}
% Text
% \end{thebibliography}
%
% or
%
% Compile your BiBTeX database using our plos2015.bst
% style file and paste the contents of your .bbl file
% here. See http://journals.plos.org/plosone/s/latex for 
% step-by-step instructions.
% 
%\begin{thebibliography}{10}
%
%\bibitem{bib1}
%Conant GC, Wolfe KH.
%\newblock {{T}urning a hobby into a job: how duplicated genes find new
%  functions}.
%\newblock Nat Rev Genet. 2008 Dec;9(12):938--950.
%
%\bibitem{bib2}
%Ohno S.
%\newblock Evolution by gene duplication.
%\newblock London: George Alien \& Unwin Ltd. Berlin, Heidelberg and New York:
%  Springer-Verlag.; 1970.
%
%\bibitem{bib3}
%Magwire MM, Bayer F, Webster CL, Cao C, Jiggins FM.
%\newblock {{S}uccessive increases in the resistance of {D}rosophila to viral
%  infection through a transposon insertion followed by a {D}uplication}.
%\newblock PLoS Genet. 2011 Oct;7(10):e1002337.
%
%\end{thebibliography}

\end{document}